\documentclass[12]{article}

\usepackage{makeidx}         
\usepackage{graphicx}        
\usepackage{multicol}        
\usepackage[bottom]{footmisc}

\usepackage{supertabular}
\usepackage{fancyhdr}
\makeindex             

\begin{document}

\title{Study on Delaunay tessellations of 1-irregular cuboids for 3D mixed element meshes}
\author{David Contreras and Nancy Hitschfeld-Kahler \\
Department of Computer Science,
        FCFM, University of Chile, Chile \\
E-mails:~dcontrer,nancy@dcc.uchile.cl}
%
%
\date{}
\maketitle

\begin{abstract}
Mixed elements meshes based on the modified octree approach contain
several co-spherical point configurations. While generating Delaunay
tessellations to be used together with the finite volume method, it is
not necessary to partition them into tetrahedra; co-spherical elements
can be used as final elements. This paper presents
a study  of all co-spherical elements  that  appear while tessellating
a 1-irregular cuboid (cuboid with at most one Steiner point on its edges) with
different aspect ratio. Steiner points can be located at any position between the 
edge endpoints.  When Steiner points are located at edge midpoints,
24 co-spherical elements appear while tessellating 1-irregular cubes.
By inserting internal faces and edges to these new elements, this number
is reduced to 13. When 1-irregular cuboids with aspect ratio 
equal to $\sqrt{2}$ are tessellated, 10 co-spherical elements are 
required. If 1-irregular cuboids have  aspect ratio between
1 and $\sqrt{2}$, all the tessellations are adequate for the finite volume
method.  When Steiner points are located at any position, the
study was done for a specific Steiner point distribution on a cube. 38
co-spherical elements  were required to tessellate all the generated 1-irregular cubes. 
Statistics about the impact of each new element in the tessellations of
1-irregular cuboids are also included.
This study was done by developing an algorithm that 
construct  Delaunay
tessellations by starting from a Delaunay tetrahedral mesh built by Qhull. 
\end{abstract}

\section{Introduction}
\label{sec:intro}

\begin{figure}
  \centering
  \includegraphics[scale=0.5]{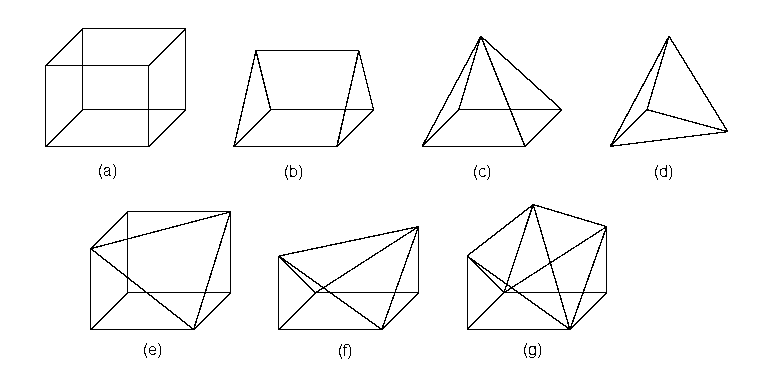}
  \caption[Ejemplo de malla mixta sobre un cuboide y su diagrama de Voronoi]%
  {The seven final elements of the $\Omega$ Mesh Generator: (a) Cuboid, (b) Triangular Prism, (c) Quadrilateral Pyramid, (d) Tetrahedron, (e) Tetrahedron Complement, (f) Deformed Prism, and (g) Deformed Tetrahedron Complement.}
  \label{fig:seven}
\end{figure}

Scientific and engineering problems are usually modeled by a set of partial 
differential equations and the solution to these  partial differential equations 
is calculated through the use of numerical methods. In order to get good results, 
the object being modeled (domain)  must be discretized in a proper way 
respecting the requirements imposed by the used numerical
method. The  discretization (mesh) is  usually composed of simple cells (basic elements) 
that must represent the domain in the best possible way. 
In particular, we are interested in meshes for the finite volume method~\cite{Bank_Rose_Fichtner_83} 
which are formed by polygons (in a 2D domain) or polyhedra (in a 3D domain), 
that satisfy the Delaunay condition:  the circumcircle in 2D, or circumsphere 
in 3D, of each element does not contain any other mesh point in its interior~\cite{Delaunay}. 
The Delaunay condition is required because we use its dual structure, the Voronoi 
diagram, to model the control volumes in order  to compute an  approximated 
solutions. The basic elements used so far are triangles and quadrilaterals in 
2D, and tetrahedra, cuboids, prisms and pyramids in 3D. Meshes composed of different
elements types are called mixed element meshes~\cite{Hitschfeld92}.\\
Mixed element meshes are built on 2D or 3D domains described by 
sets of points, polygons or polyhedra depending on the application.  
We have developed a mixed element mesh generator~\cite{Hitschfeld2004} based on an extension of
octrees~\cite{Yerry_Shephard,Shephard_90}.  Our  approach starts enclosing the domain in the smallest bounding box (cuboid). Second,
this cuboid is continuously
refined, at any edge position, by using the geometry information of the
domain. That is why this refinement is called intersection based approach. Once this step
finishes, an initial non-conforming mesh composed of  tetrahedra, pyramids, prisms, and cuboids is generated
that fits the domain geometry.
Third, these elements are further refined by bisection, as far as possible,  
until the density requirements are fulfilled.  Fourth,
the mesh is done 1-irregular by allowing only one Steiner point on each edge.
The current solution is
based on patterns but only the most frequently used patterns are available. Then, if
a pattern is not available or the element can not be properly tessellated for the
finite volume method, new Steiner points are inserted until all 1-irregular 
elements can be properly tessellated. 
The current set of seven final elements is shown in Figure~\ref{fig:seven}.

\begin{figure}
  \centering
  \includegraphics[scale=0.30]{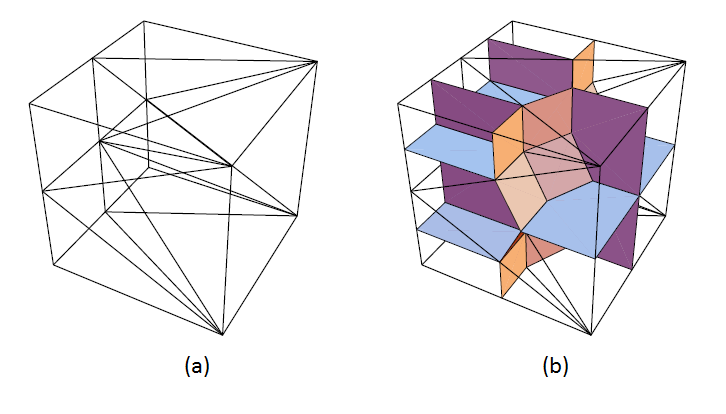}
  \caption[Ejemplo de malla mixta sobre un cuboide y su diagrama de Voronoi]%
  {(a) Mixed mesh of a 1-irregular cuboid that satisfies the Delaunay condition, (b) the same mixed mesh and its associated Voronoi diagram.}
  \label{fig:vor}
\end{figure}

The advantage of using a mixed mesh in comparison with a tetrahedral mesh is that 
the use of different element types reduce the amount of edges, faces and elements 
in the final mesh. For example, we do not need to divide a cuboid into tetrahedra. 
On the other hand, a disadvantage is that the equations must be discretized 
using different elements.
\\  Octree based approaches naturally produces co-spherical point sets. 
A  mixed mesh satisfying the Delaunay condition can include 
all produced co-spherical elements as shown in Figure~\ref{fig:vor}.
The final  elements in this example are five pyramids and four tetrahedra. 
\\The goal of this paper is to study the co-spherical elements that can appear while
tessellating 1-irregular cuboids generated by using a bisection and intersection 
based approach and to analyze how
useful would be to include the new elements in the final element set. 
In particular, this paper gives the number and shape of new co-spherical 
elements needed to tessellate (a) all 1-irregular cuboids generated by the 
bisection approach and  (b)  some particular 1-irregular cuboids generated by the
intersection refinement approach. In addition, statistics 
associated with particular tessellations are presented such as the frequency each co-spherical element
is used and the number of tessellations that can be used with the   
finite volume method without adding extra vertices. 
The analysis of the 1-irregular cuboid tessellations was done under different criteria that affect the amount of generated co-spherical elements.
\\ We have focused this work on the analysis of the tessellations of 1-irregular 
cuboids because this element
is the one that more frequently appears when meshes are generated by a 
modified octree approach. A theoretical study on the number of different 1-irregular 
cuboid configurations that can appear either by using a bisection or an intersection 
based approach was published in~\cite{Hitschfeld2000b}. We use the results of that work as starting point for this study.   
\\This paper is organized as follows: Section~\ref{section2}
describes the bisection and intersection refinement approaches. 
Section~\ref{algo} presents briefly the developed algorithm to compute
Delaunay tessellations. Section~\ref{Results} and Section~\ref{ResultsI} give the results
obtained by applying the algorithm to 1-irregular cuboids generated
by a bisection and an intersection based approach, respectively. Section~\ref{conclusiones} includes our conclusions.

\section{Basic concepts}
\label{section2}
This section describes some ideas in order to understand 
how 1-irregular cuboids are generated.

\subsection{Bisection based approach 1-irregular configurations}\label{bisec}

\begin{figure}
  \centering
  \includegraphics[scale=0.4]{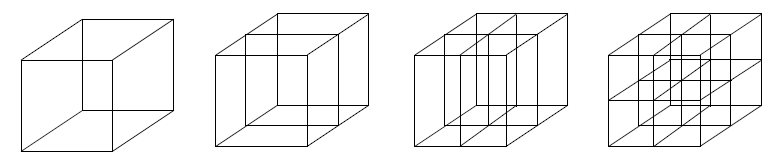}
  \caption[División de un cuboide mediante un refinado por bisección de aristas]%
  {Cuboid and its splits into two, four and eight cuboids using a bisection based approach.}
  \label{fig:cbis}
\end{figure}

\begin{figure}
  \centering
  \includegraphics[scale=0.45]{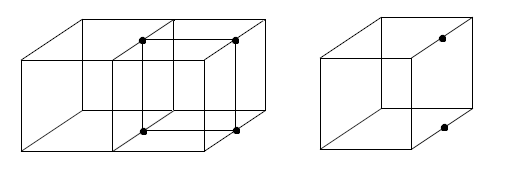}
  \caption[Cuboide 1-irregular generado mediante un refinado por bisección de aristas]%
  {The bisection-refined cuboid at the left produces an 1-irregular element like the cuboid 
at the right}.
  \label{fig:cbisref}
\end{figure}

In this approach, the Steiner points inserted at the refinement phase are always located at the edge midpoints. Cuboids can be refined into two, four or eight smaller cuboids as shown in Figure~\ref{fig:cbis}.
\\This refinement produces neighboring cuboids with Steiner points located at the edge midpoints. Those 1-irregular cuboids are larger than the already refined neighbor cuboid as shown in Figure~\ref{fig:cbisref}.

\subsection{Intersection based approach 1-irregular configurations}\label{inter}
\begin{figure}
  \centering
  \includegraphics[scale=0.4]{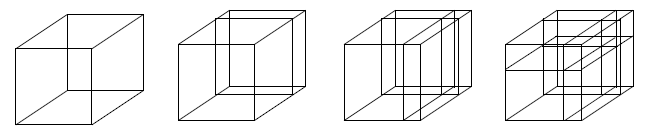}
  \caption[División de un cuboide mediante un refinado por intersección]%
  {Cuboid and its splits into two, four and eight cuboids using an intersection based approach.}
  \label{fig:cint}
\end{figure}

While using an intersection based approach, the Steiner points are not necessarily 
located at the edge midpoints. In general, there are no constrains on the location 
of the Steiner points, except by the fact that parallel edges must be divided in the same relative position to ensure the generation of cuboids and not any other polyhedron. Figure~\ref{fig:cint} shows an example of this approach.
\\This refinement  produces neighboring cuboids with Steiner points located at any edge position. Those 1-irregular cuboids are larger than the already refined neighbor cuboid as shown in Figure~\ref{fig:cbisref}.
\begin{figure}
  \centering
  \includegraphics[scale=0.5]{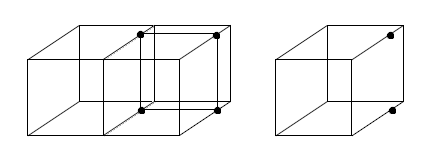}
  \caption[Cuboide 1-irregular generado mediante un refinado por intersección]%
  {The intersection-refined cuboid to the left produces an 1-irregular element like the cuboid to the right.}
  \label{fig:cintref}
\end{figure}

\section{Algorithm}\label{algo}

In order to count the number of new co-spherical elements than can appear and to
recognize their shape, we have developed an algorithm that executes the followings steps:\\
\begin{enumerate}
\item Build the point configuration of a 1-irregular cuboid by specifying the coordinates 
of the cuboid vertices and its Steiner points.
\item Build a Delaunay tetrahedral mesh for this point configuration by using QHull~\cite{Barber96thequickhull}\footnote{http://www.qhull.org}.
\item Join tetrahedra to form the largest possible co-spherical elements.
\item Identify each final co-spherical polyhedron.
\end{enumerate}
Qhull divides co-spherical point configurations into a set of tetrahedra by adding an
artificial point that is not part of the input. Then, we use this fact to
recognize the faces that form a co-spherical polyhedron and later to recognize
which element is.

\section{Results: Bisection based approach}\label{Results}

This section describes the results obtained by applying the previous
algorithm to the 4096 ($2^{12}$) 1-irregular configurations that can 
be generated using a bisection based approach. First, the new 
co-spherical elements are shown. Then, their impact in all the
tessellations is analyzed and finally, the tessellations that
can be used with the finite volume method are characterized.  
\subsection{New co-spherical elements}

We have identified 17 
new co-spherical polyhedra in the tessellations of 1-irregular cubes 
in addition to the seven original elements shown 
in Figure~\ref{fig:seven}. 
A description of each one can be 
found in Table~\ref{tab:bisecelem}. A distinction is made between rectangular 
and quadrilateral faces except for the quadrilateral pyramid. 
Because of this, the triangular prism and the generic element \#1 are considered 
different  co-spherical elements, and the same happens between the 
deformed prism and the generic element \# 3. This could be changed in a future
study.\\

\begin{center}
\tablefirsthead{%
\hline
\multicolumn{1}{|c@{\hspace{5mm}}}{{\bf Element}} &
\multicolumn{1}{|c@{\hspace{5mm}}}{{\bf Vertices}} &
\multicolumn{1}{|c@{\hspace{5mm}}}{{\bf Edges}} &
\multicolumn{1}{|c@{\hspace{5mm}}}{{\bf Faces}} &
\multicolumn{1}{|c@{\hspace{5.5mm}}|}{{\bf Example}} \\
\hline}
\tablehead{%
\hline
\multicolumn{5}{|l|}{\small\sl continued from previous page}\\
\hline
\multicolumn{1}{|c@{\hspace{5mm}}}{{\bf Element}} &
\multicolumn{1}{|c@{\hspace{5mm}}}{{\bf Vertices}} &
\multicolumn{1}{|c@{\hspace{5mm}}}{{\bf Edges}} &
\multicolumn{1}{|c@{\hspace{5mm}}}{{\bf Faces}} &
\multicolumn{1}{|c@{\hspace{5.5mm}}|}{{\bf Example}} \\
\hline}
\tabletail{%
\hline
\multicolumn{5}{|r|}{\small\sl continued on next page}\\
\hline}
\tablelasttail{\hline}
\topcaption{Description of the new co-spherical elements that appear while
tessellating 1-irregular cubes}
\begin{supertabular}{| c@{\hspace{4mm}} | c@{\hspace{5.5mm}} | c@{\hspace{5mm}} | c@{\hspace{5mm}} | c@{\hspace{5.5mm}} |}
\shrinkheight{85.0mm}
\label{tab:bisecelem}
Pentagonal Pyramid &  6 & 10 & 6 & \raisebox{-\totalheight}{\includegraphics[scale=0.125]{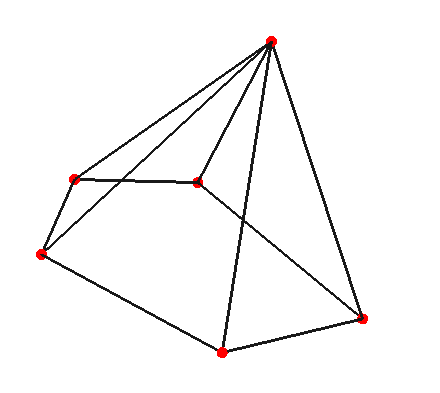}} \\ \hline
Hexagonal Pyramid &  7 & 12 & 7 & \raisebox{-\totalheight}{\includegraphics[scale=0.125]{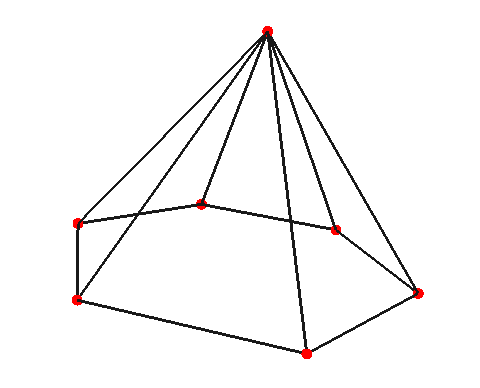}} \\ \hline
Triangular Bipyramid &  5 & 9 & 6 & \raisebox{-\totalheight}{\includegraphics[scale=0.125]{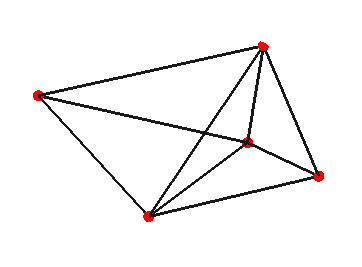}} \\ \hline
Quadrilateral Bipyramid &  6 & 12 & 8 & \raisebox{-\totalheight}{\includegraphics[scale=0.125]{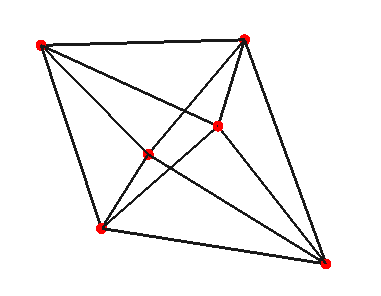}} \\ \hline
Pentagonal Bipyramid &  7 & 15 & 10 & \raisebox{-\totalheight}{\includegraphics[scale=0.125]{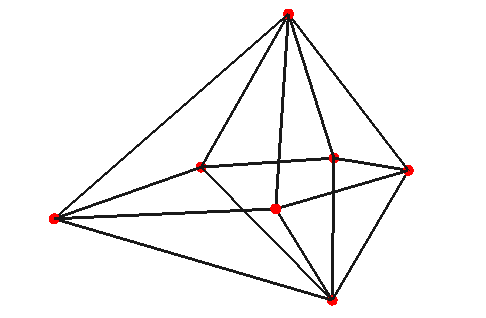}} \\ \hline
Hexagonal Bipyramid &  8 & 18 & 12 & \raisebox{-\totalheight}{\includegraphics[scale=0.125]{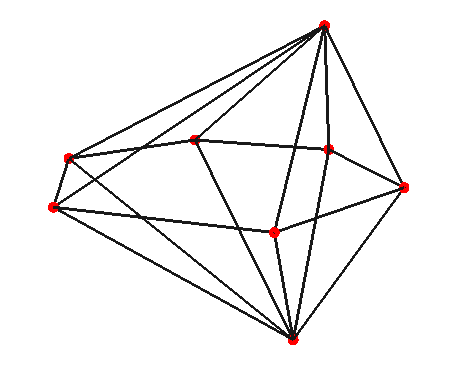}} \\ \hline
Triangular Biprism &  8 & 14 & 8 & \raisebox{-\totalheight}{\includegraphics[scale=0.125]{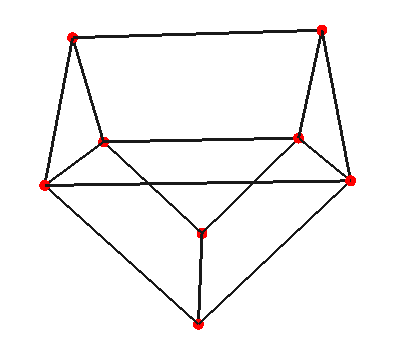}} \\ \hline
Generic \#1 & 6 & 9 & 5 & \raisebox{-\totalheight}{\includegraphics[scale=0.125]{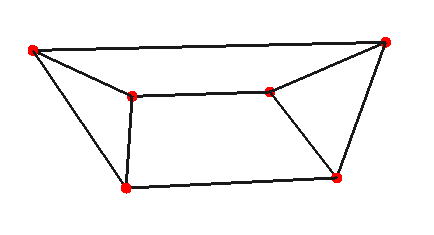}} \\ \hline
Generic \#2 & 6 & 10 & 6 & \raisebox{-\totalheight}{\includegraphics[scale=0.125]{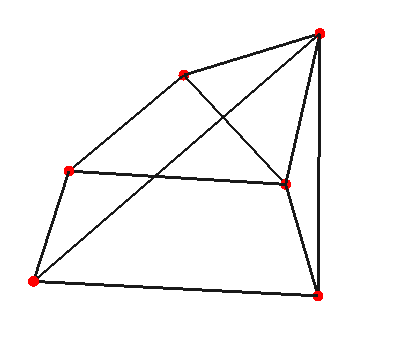}} \\ \hline
Generic \#3 & 6 & 11 & 7 & \raisebox{-\totalheight}{\includegraphics[scale=0.125]{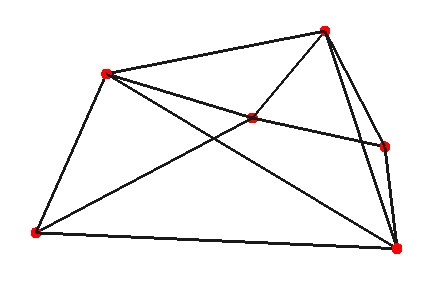}} \\ \hline
Generic \#4 &  7 & 12 & 7 & \raisebox{-\totalheight}{\includegraphics[scale=0.125]{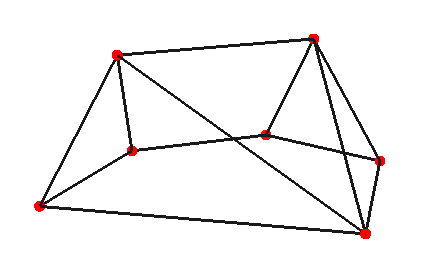}} \\ \hline
Generic \#5 & 7 & 13 & 8 & \raisebox{-\totalheight}{\includegraphics[scale=0.125]{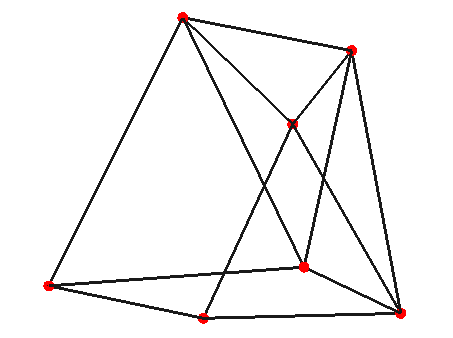}} \\ \hline
Generic \#6 & 8 & 15 & 9 & \raisebox{-\totalheight}{\includegraphics[scale=0.125]{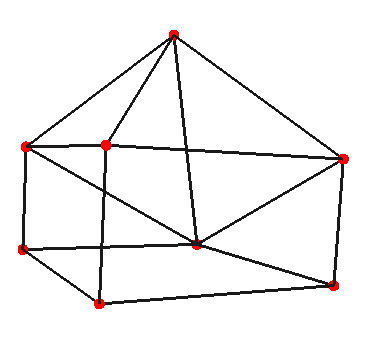}} \\ \hline
Generic \#7 & 8 & 16 & 10 & \raisebox{-\totalheight}{\includegraphics[scale=0.125]{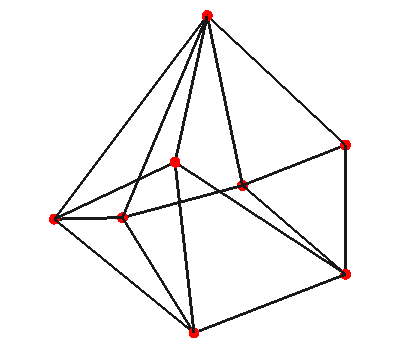}} \\ \hline
Generic \#8 & 8 & 17 & 11 & \raisebox{-\totalheight}{\includegraphics[scale=0.125]{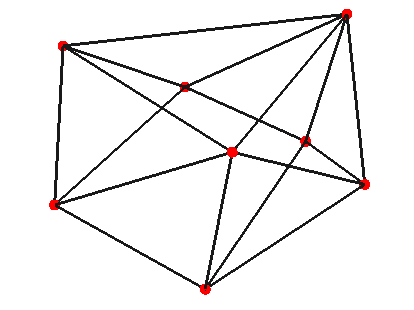}} \\ \hline
Generic \#9 & 9 & 16 & 9 & \raisebox{-\totalheight}{\includegraphics[scale=0.125]{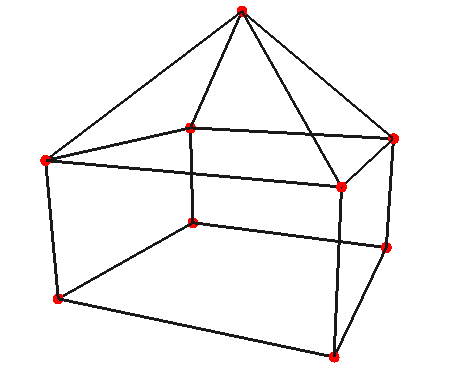}} \\ \hline
Generic \#10 & 9 & 18 & 11 & \raisebox{-\totalheight}{\includegraphics[scale=0.125]{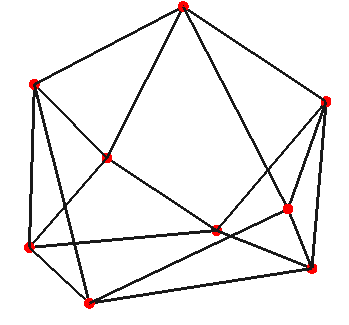}} \\ \hline
\end{supertabular} \label{table:bis}
\end{center}

\subsection{Element analysis}\label{criteria}

\begin{figure}[hu]
  \centering
  \includegraphics[scale=0.2]{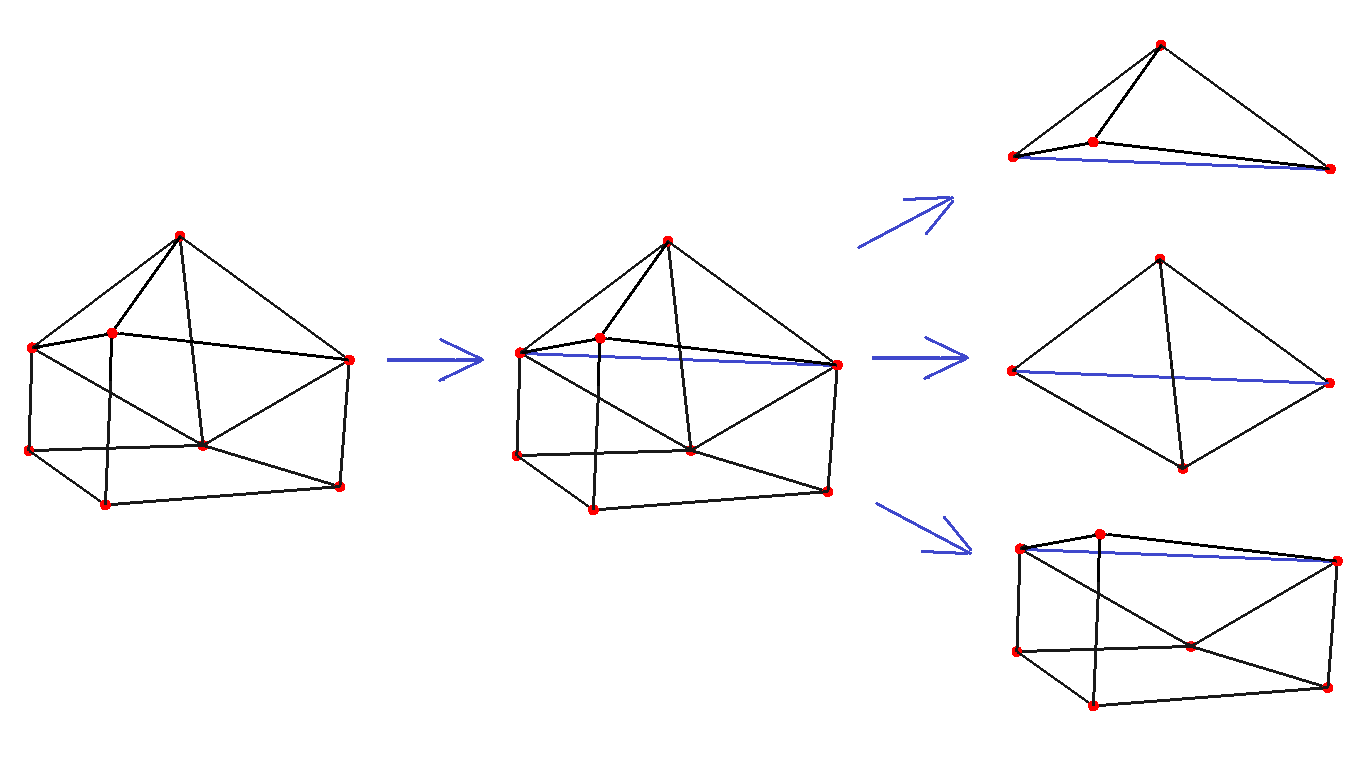}
  \caption[Ejemplo de elemento coesférico separable agregando aristas internas]%
  {Generic \#6 element and its separation into three different elements by adding an extra inner edge shown in blue.}
  \label{fig:SepComplex}
\end{figure}

Since there are 17 new co-spherical elements, the natural question is if this
number can be reduced without adding diagonals in the cuboid rectangular faces. In
fact, our mixed element mesh generator requires to tessellate 1-irregular cuboids without
adding diagonals on its rectangular faces when it uses a pattern-wise approach.  In the following, we analyze
the number of co-spherical elements  under three different criteria:
\begin{itemize}
\item {\bf Finding the optimal tessellation}: An optimal tessellation contains the lowest amount of final elements. This is reached by maximizing the number of elements 
with different shape. The  number of co-spherical elements that
can be used is 24.
\item {\bf Minimizing the number of different co-spherical elements by adding only internal faces}: Under this criterion we want to reduce the number of different final elements by adding only internal faces. 
Examining the set of new elements in Table~\ref{table:bis}, we see that the bipyramids and the biprisms are naturally divisible into two elements, and so are the generic \#5 (separable into a prism and a quadrilateral pyramid), generic \#8 (separable into a prism and two quadrilateral pyramids) and generic \#9 (separable into a cuboid and quadrilateral pyramid),
among others. An example of this type of separation is shown in Figure~\ref{fig:SepSimple}. The total 
number of co-spherical elements needed to tessellate the 4096 configurations is now 16.
\begin{figure}[hu]
  \centering
  \includegraphics[scale=0.25]{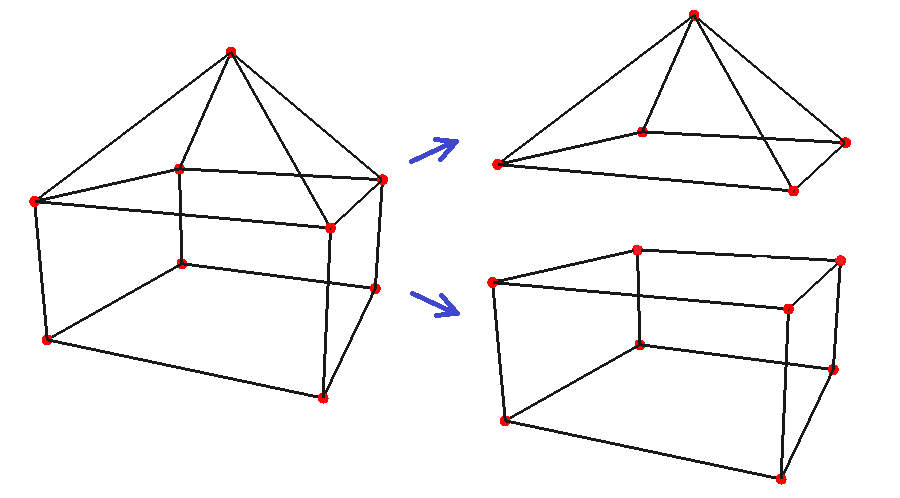}
  \caption[Ejemplo de elemento coesférico separable directamente]%
  {Generic \#9 element and its separation into two different elements.}
  \label{fig:SepSimple}
\end{figure}
\item{\bf Minimizing the number of different co-spherical elements by adding internal edges and faces}: This extends the second criterion by adding the condition that it is  possible to add extra edges only if they are inside the new elements. The reason for only allowing internal edges is that adding external edges could change the partition of one of the rectangular
faces of the original cuboid. Under this criterion, the elements that are separable are generic \#3 (one inner edge produces two tetrahedra and one quadrilateral pyramid), generic \#6 (one inner edge produces two tetrahedra and a tetrahedron complement) and generic \#7 (two inner edges produce two tetrahedra, a quadrilateral pyramid and a deformed prism). An example of this type of separation is shown in Figure~\ref{fig:SepComplex}. The total number of 
co-spherical elements needed to tessellate the 4096 configurations is reduced to 13.
\end{itemize}

\subsection{Evaluating the impact of each new element}

In order to decide how important is to include a new element in the final element set, 
in this section we study how many times each co-spherical element appears
in the tessellation of a 1-irregular cuboid. For this study, we have run our program for
1-irregular cuboids with three different aspect ratio: 1, 4, and $\sqrt{2}$. 

\begin{itemize}
\item
 {\bf Test case A}. Aspect ratio equal to 1 ($a=b=c$): The 1-irregular cube appears naturally on the standard octree and this method is used by most mesh generators based on octrees. 

\item 
{\bf Test case B}. Aspect ratio equal to 4 ($4a=2b=c$):  This represents a typical cuboid
to model thin zones.

\item 
{\bf Test case C}. Aspect ratio equal to $\sqrt{2}$ ($a\sqrt{2}=b=c$): It was shown in~\cite{Conti_diss} that some 1-irregular cuboid within these proportions can be tessellated without problems for the finite volume method.

\end{itemize}


\subsubsection{Running the test case A}

Table~\ref{tabla2} shows the frequency in which appear each one of the 24 
co-spherical elements in the tessellations of  1-irregular cubes. \\ \bigskip 
\begin{center}
\tablefirsthead{%
\hline
\multicolumn{1}{|c@{\hspace{5mm}}}{{\bf Element}} &
\multicolumn{1}{|c@{\hspace{5.5mm}}||}{{\bf Freq.}} &
\multicolumn{1}{|c@{\hspace{5mm}}}{{\bf Element}} &
\multicolumn{1}{|c@{\hspace{5.5mm}}|}{{\bf Freq.}} \\
\hline}
\tablehead{%
\hline
\multicolumn{4}{|l|}{\small\sl continued from previous page}\\
\hline
\multicolumn{1}{|c@{\hspace{5mm}}}{{\bf Element}} &
\multicolumn{1}{|c@{\hspace{5.5mm}}||}{{\bf Freq.}} &
\multicolumn{1}{|c@{\hspace{5mm}}}{{\bf Element}} &
\multicolumn{1}{|c@{\hspace{5.5mm}}|}{{\bf Freq.}} \\
\hline}
\tabletail{%
\hline
\multicolumn{4}{|r|}{\small\sl continued on next page}\\
\hline}
\tablelasttail{}
\topcaption{Frequency of the co-spherical elements on 1-irregular cube
tessellations} \label{tabla2}
\begin{supertabular}{| l@{\hspace{5mm}} | r@{\hspace{5.5mm}} || l@{\hspace{5mm}} | r@{\hspace{5.5mm}} | }
Cuboid & 195 & 				Hexagonal Bipyramid & 36\\
Tetrahedron & 18,450 &			Triangular Biprism & 6\\
Quadrilateral Pyramid & 11,718 & 	Generic \#1 & 12\\
Triangular Prism & 3,720 &	 		Generic \#2 & 96\\
Tetrahedron Comp. & 992 & 		Generic \#3 & 48\\
Def. Prism & 396 & 				Generic \#4 & 48\\
Def. Tetrahedron Comp. & 144 &		Generic \#5 & 120\\
Pentagonal Pyramid & 384 &		Generic \#6 & 24\\
Hexagonal Pyramid & 56 &			Generic \#7 & 48\\
Triangular Bipyramid & 240 & 		Generic \#8 & 48\\
Quadrilateral Bipyramid & 272 &		Generic \#9  & 6\\
Pentagonal Bipyramid & 192 &		Generic \#10 & 8 \\ \hline
\multicolumn{1}{c}{} & \multicolumn{1}{c}{} & \multicolumn{1}{|l}{{\bf Total}} & \multicolumn{1}{|r|}{{\bf 37,259}}  \\
\cline{3-4}
\end{supertabular}
\end{center}
From Table~\ref{tabla2}, we observe that the most used elements correspond 
to tetrahedra and quadrilateral pyramids ($\sim$49.5\% and $\sim$31.5\% of the total elements, respectively). Moreover, the set of seven initial co-spherical elements represents  $\sim$95.6\% of the total. If the number of co-spherical elements is reduced to 16 by
adding internal faces, the element frequencies are distributed as shown in
Table~\ref{tabla3}. It can be observed that
the most used elements are again tetrahedra 
and quadrilateral pyramids ($\sim$49.5\% and $\sim$32.7\% of elements, respectively). 
The set of seven initial co-spherical elements represents  $\sim$96.9\% of the total.\\ \bigskip
\begin{center}
\tablefirsthead{%
\hline
\multicolumn{1}{|c@{\hspace{5mm}}}{{\bf Element}} &
\multicolumn{1}{|c@{\hspace{5.5mm}}||}{{\bf Freq.}} &
\multicolumn{1}{|c@{\hspace{5mm}}}{{\bf Element}} &
\multicolumn{1}{|c@{\hspace{5.5mm}}|}{{\bf Freq.}} \\
\hline}
\tablehead{%
\hline
\multicolumn{4}{|l|}{\small\sl continued from previous page}\\
\hline
\multicolumn{1}{|c@{\hspace{5mm}}}{{\bf Element}} &
\multicolumn{1}{|c@{\hspace{5.5mm}}||}{{\bf Freq.}} &
\multicolumn{1}{|c@{\hspace{5mm}}}{{\bf Element}} &
\multicolumn{1}{|c@{\hspace{5.5mm}}|}{{\bf Freq.}} \\
\hline}
\tabletail{%
\hline
\multicolumn{4}{|r|}{\small\sl continued on next page}\\
\hline}
\tablelasttail{}
\topcaption{Frequency of the co-spherical elements adding only internal faces on 1-irregular cube tessellations}
\begin{supertabular}{| l@{\hspace{5mm}} | r@{\hspace{5.5mm}} || l@{\hspace{5mm}} | r@{\hspace{5.5mm}} | }
     Cuboid & 201 &			 	Hexagonal Pyramid & 128\\
     Tetrahedron & 18,930 &	 	Generic \#1 & 12\\
     Quadrilateral Pyramid & 12,484 &	Generic \#2 & 96\\
     Triangular Prism & 3,900 &     		Generic \#3 & 48\\
     Tetrahedron Comp. & 992 &     	Generic \#4 & 48\\
     Def. Prism & 396 &     			Generic \#6 & 24\\
     Def. Tetrahedron Comp. & 144 &     	Generic \#7 & 48\\
     Pentagonal Pyramid & 768 &     	Generic \#10 & 8 \\ \hline
\multicolumn{1}{c}{} & \multicolumn{1}{c}{} & \multicolumn{1}{|l}{{\bf Total}} & \multicolumn{1}{|r|}{{\bf 38,227}}  \\
\cline{3-4}
\end{supertabular}\label{tabla3}
\end{center}
Finally, when the number of different final co-spherical elements is reduced to 13, by
adding internal edges and faces, the element frequencies are shown in Table~\ref{tabla4}.
The most used elements correspond to tetrahedra and quadrilateral pyramids ($\sim$49.8\% and $\sim$32.7\% of the total number of elements, respectively). The set of  initial seven co-spherical elements represents  $\sim$97.2\% of the total.
\\ \bigskip
\begin{center}
\tablefirsthead{%
\hline
\multicolumn{1}{|c@{\hspace{5mm}}}{{\bf Element}} &
\multicolumn{1}{|c@{\hspace{5.5mm}}||}{{\bf Freq.}} &
\multicolumn{1}{|c@{\hspace{5mm}}}{{\bf Element}} &
\multicolumn{1}{|c@{\hspace{5.5mm}}|}{{\bf Freq.}} \\
\hline}
\tablehead{%
\hline
\multicolumn{4}{|l|}{\small\sl continued from previous page}\\
\hline
\multicolumn{1}{|c@{\hspace{5mm}}}{{\bf Element}} &
\multicolumn{1}{|c@{\hspace{5.5mm}}||}{{\bf Freq.}} &
\multicolumn{1}{|c@{\hspace{5mm}}}{{\bf Element}} &
\multicolumn{1}{|c@{\hspace{5.5mm}}|}{{\bf Freq.}} \\
\hline}
\tabletail{%
\hline
\multicolumn{4}{|r|}{\small\sl continued on next page}\\
\hline}
\tablelasttail{}
\topcaption{Frequency of co-spherical elements adding internal faces and edges on 1-irregular cube tessellations}
\begin{supertabular}{| l@{\hspace{5mm}} | r@{\hspace{5.5mm}} || l@{\hspace{5mm}} | r@{\hspace{5.5mm}} | }
     Cuboid & 201 &				Pentagonal Pyramid & 768\\
     Tetrahedron & 19,170 &		Hexagonal Pyramid & 128\\
     Quadrilateral Pyramid & 12,580 &     	Generic \#1 & 12\\
     Triangular Prism & 3,900 &		Generic \#2 & 96\\
     Tetrahedron Comp. & 1,016 &    	Generic \#4 & 48\\
     Def. Prism & 444 &     			Generic \#10 & 8 \\
     Def. Tetrahedron Comp. & 144 & & \\ \hline
\multicolumn{1}{c}{} & \multicolumn{1}{c}{} & \multicolumn{1}{|l}{{\bf Total}} & \multicolumn{1}{|r|}{{\bf 38,515}}  \\
\cline{3-4}
\end{supertabular} \label{tabla4}
\end{center}

\subsubsection{Running the test Case B}

When the aspect ratio of the cuboid is changed to 4, only 6 different  co-spherical elements appear and their frequencies are shown in Table~\ref{tabla5}.\\ 
\begin{center}{t}
\tablefirsthead{%
\hline
\multicolumn{1}{|c@{\hspace{5mm}}}{{\bf Element}} &
\multicolumn{1}{|c@{\hspace{5.5mm}}||}{{\bf Freq.}} &
\multicolumn{1}{|c@{\hspace{5mm}}}{{\bf Element}} &
\multicolumn{1}{|c@{\hspace{5.5mm}}|}{{\bf Freq.}} \\
\hline}
\tablehead{%
\hline
\multicolumn{4}{|l|}{\small\sl continued from previous page}\\
\hline
\multicolumn{1}{|c@{\hspace{5mm}}}{{\bf Element}} &
\multicolumn{1}{|c@{\hspace{5.5mm}}||}{{\bf Freq.}} &
\multicolumn{1}{|c@{\hspace{5mm}}}{{\bf Element}} &
\multicolumn{1}{|c@{\hspace{5.5mm}}|}{{\bf Freq.}} \\
\hline}
\tabletail{%
\hline
\multicolumn{4}{|r|}{\small\sl continued on next page}\\
\hline}
\tablelasttail{}
\topcaption{Frequency of the co-spherical elements in the tessellations of  1-irregular cuboids with aspect ratio 4}
\begin{supertabular}{| l@{\hspace{5mm}} | r@{\hspace{5.5mm}} || l@{\hspace{5mm}} | r@{\hspace{5.5mm}} | }
     Cuboid & 103 &				Triangular Prism & 3,120\\
     Tetrahedron & 29,118 &     		Tetrahedron Comp. & 536\\
     Quadrilateral Pyramid & 12,620 &     	Def. Prism & 84\\ \hline
\multicolumn{1}{c}{} & \multicolumn{1}{c}{} & \multicolumn{1}{|l}{{\bf Total}} & \multicolumn{1}{|r|}{{\bf 45,581}}  \\
\cline{3-4}
\end{supertabular} \label{tabla5}
\end{center}
From Table~\ref{tabla5}, we observe that tetrahedra and quadrilateral pyramids are the most
used elements, comprising more than 90\% of the total of the elements ($\sim$63.9\% of tetrahedra and $\sim$27.7\% quadrilateral pyramids).  Note that  these elements can not be divided
into simpler ones without adding diagonals on its quadrilateral faces. 

\subsubsection{Running the test Case C}
When the aspect ratio is equal to $\sqrt{2}$, only 
10 different final co-spherical elements appear whose frequencies are distributed as follows:\\ \bigskip
\begin{center}
\tablefirsthead{%
\hline
\multicolumn{1}{|c@{\hspace{5mm}}}{{\bf Element}} &
\multicolumn{1}{|c@{\hspace{5.5mm}}||}{{\bf Freq.}} &
\multicolumn{1}{|c@{\hspace{5mm}}}{{\bf Element}} &
\multicolumn{1}{|c@{\hspace{5.5mm}}|}{{\bf Freq.}} \\
\hline}
\tablehead{%
\hline
\multicolumn{4}{|l|}{\small\sl continued from previous page}\\
\hline
\multicolumn{1}{|c@{\hspace{5mm}}}{{\bf Element}} &
\multicolumn{1}{|c@{\hspace{5.5mm}}||}{{\bf Freq.}} &
\multicolumn{1}{|c@{\hspace{5mm}}}{{\bf Element}} &
\multicolumn{1}{|c@{\hspace{5.5mm}}|}{{\bf Freq.}} \\
\hline}
\tabletail{%
\hline
\multicolumn{4}{|r|}{\small\sl continued on next page}\\
\hline}
\tablelasttail{}
\topcaption{Frequency of the final elements on optimal tessellations on 1-irregular cuboids with aspect ratio $\sqrt{2}$}
\begin{supertabular}{| l@{\hspace{5mm}} | r@{\hspace{5.5mm}} || l@{\hspace{5mm}} | r@{\hspace{5.5mm}} | }
\shrinkheight{-5.0mm}
     Cuboid & 199 &				Def. Prism & 284\\
     Tetrahedron & 25,252 &   		Triangular Bipyramid  & 128\\
     Quadrilateral Pyramid & 12,300 &     	Quadrilateral Bipyramid & 52\\
     Triangular Prism & 3,780 &     		Generic \#2 & 16\\
     Tetrahedron Comp. & 1,008 &     	Generic \#5 & 16\\ \hline
\multicolumn{1}{c}{} & \multicolumn{1}{c}{} & \multicolumn{1}{|l}{{\bf Total}} & \multicolumn{1}{|r|}{{\bf 43,035}}  \\
\cline{3-4}
\end{supertabular}
\end{center}
The most used elements correspond to tetrahedra and quadrilateral pyramids ($\sim$58.7\% and $\sim$28.6\% of the total number of elements, respectively). Moreover, the set of initial seven co-spherical elements represents a $\sim$99.5\% of the total. In this test case, the number of co-spherical elements can be
reduced from 10 to 7 by adding internal faces.  The frequencies of
these seven elements are distributed as shown in Table~\ref{tablaR2}.\\ \bigskip
\begin{center}
\tablefirsthead{%
\hline
\multicolumn{1}{|c@{\hspace{5mm}}}{{\bf Element}} &
\multicolumn{1}{|c@{\hspace{5.5mm}}||}{{\bf Freq.}} &
\multicolumn{1}{|c@{\hspace{5mm}}}{{\bf Element}} &
\multicolumn{1}{|c@{\hspace{5.5mm}}|}{{\bf Freq.}} \\
\hline}
\tablehead{%
\hline
\multicolumn{4}{|l|}{\small\sl continued from previous page}\\
\hline
\multicolumn{1}{|c@{\hspace{5mm}}}{{\bf Element}} &
\multicolumn{1}{|c@{\hspace{5.5mm}}||}{{\bf Freq.}} &
\multicolumn{1}{|c@{\hspace{5mm}}}{{\bf Element}} &
\multicolumn{1}{|c@{\hspace{5.5mm}}|}{{\bf Freq.}} \\
\hline}
\tabletail{%
\hline
\multicolumn{4}{|r|}{\small\sl continued on next page}\\
\hline}
\tablelasttail{}
\topcaption{Frequency of the 7 co-spherical  elements while tessellating
1-irregular cuboids with aspect ratio $\sqrt{2}$}
\begin{supertabular}{| l@{\hspace{5mm}} | r@{\hspace{5.5mm}} || l@{\hspace{5mm}} | r@{\hspace{5.5mm}} | }
     Cuboid & 199 &			     	Tetrahedron Comp. & 1,008\\
     Tetrahedron & 25,508 &		Def. Prism & 284\\
     Quadrilateral Pyramid & 12,420 &     	Generic \#2 & 16\\
     Triangular Prism & 3,796 & & \\ \hline
\multicolumn{1}{c}{} & \multicolumn{1}{c}{} & \multicolumn{1}{|l}{{\bf Total}} & \multicolumn{1}{|r|}{{\bf 43,231}}  \\
\cline{3-4}
\end{supertabular}\label{tablaR2}
\end{center}
Again, the most used elements are tetrahedra and quadrilateral pyramids ($\sim$59.0\% and $\sim$28.7\% of elements, respectively). There are only 6 of the seven initial co-spherical elements, representing a $\sim$99.96\% of the total.
Notice that this set of 7 elements is not separable by adding internal
edges or faces.

\subsection{Tessellations and the finite volume method}\label{Calidad}

We have also examined whether the generated tessellations meet the 
requirements for their use in the context of the finite volume method.
The requirement is  that the circumcenter of each final element is 
contained within the initial 1-irregular cuboid.
This requirement is strong but it allows our mesh generator to find a proper tessellation of each 1-irregular cuboid locally.\\
\noindent The evaluation of each tessellation is performed on the same test cases discussed 
in Section~\ref{criteria}. The results are shown in Table~\ref{tablaQ}.
We observe that the circumcenters of all elements are inside the initial 
cuboid for all the  configurations in the test cases  A and C. This means
that all 1-irregular configurations could be properly tessellated if the
aspect ratio of the elements is less or equal to $\sqrt{2}$. If 1-irregular
cuboids has an aspect ratio equal to 4, only 132 1-irregular cuboids 
fit the circumcenter requirement.
\begin{table}[h]
  \caption{Number of configurations that fit the circumcenter requirement}
\begin{center}
\begin{tabular}{|c|c|}\hline
  & Number of proper configurations \\ \hline 
    Test Case A &  \\ 
    (Aspect ratio equal to 1) &  4096 \\ \hline  
    Test Case B &  \\ 
    (Aspect ratio equal to  4) &  132 \\ \hline  
    Test Case C &  \\ 
    (Aspect ratio equal to  $\sqrt{2}$) &  4096 \\ \hline  
  \end{tabular}\label{tablaQ}
\end{center}
\end{table}

\section{Results: Intersection based approach}\label{ResultsI}

The number of 1-irregular configurations that can appear while
refining cuboids by an intersection based approach is  $187^3+1$~\cite{Hitschfeld2000b}. We consider that two 1-irregular configurations are different
if the relative position of Steiner points located on 
parallel cuboid edges is  not the same. In this section we describe
the results obtained by applying the
algorithm to all 1-irregular configurations of a cube that can be generated by
inserting Steiner points only on the positions defined by
multiples of 1/16 of the edge length. The impact of each co-spherical
element was only obtained for the 1-irregular cube.

\subsection{New co-spherical elements}

Since the possible positions of Steiner points on a particular
edge are infinite, we only use a set of predetermined  Steiner
point positions  for each set of cuboid parallel edges
defined as follows: \\
\begin{itemize}
\item The first vertex is always located at the midpoint of an edge.
\item If the $ k $-th point is located to the left of the $ k-1$ previous 
points, its actual position is located at the midpoint of the segment 
defined by the left edge corner and the leftmost already assigned Steiner 
point. 
Similarly, if the $k$-th point is located to the right, its actual 
position is determined by the midpoint of the segment defined by 
the rightmost assigned Steiner point and the right edge corner.
\item If the relative position of the $ k $-th point is between two 
Steiner points 
already allocated, its actual position is determined by the midpoint of the two Steiner points.
\end{itemize} 
Under this approach we identified 14 new co-spherical elements in the tessellations of  1-irregular cubes. A description of each of them can be found in Table~\ref{tab:interelem}.\bigskip

\begin{center}
\tablefirsthead{%
\hline
\multicolumn{1}{|c@{\hspace{5mm}}}{{\bf Element}} &
\multicolumn{1}{|c@{\hspace{5mm}}}{{\bf Vertices}} &
\multicolumn{1}{|c@{\hspace{5mm}}}{{\bf Edges}} &
\multicolumn{1}{|c@{\hspace{5mm}}}{{\bf Faces}} &
\multicolumn{1}{|c@{\hspace{5.5mm}}|}{{\bf Example}} \\
\hline}
\tablehead{%
\hline
\multicolumn{5}{|l|}{\small\sl continued from previous page}\\
\hline
\multicolumn{1}{|c@{\hspace{5mm}}}{{\bf Element}} &
\multicolumn{1}{|c@{\hspace{5mm}}}{{\bf Vertices}} &
\multicolumn{1}{|c@{\hspace{5mm}}}{{\bf Edges}} &
\multicolumn{1}{|c@{\hspace{5mm}}}{{\bf Faces}} &
\multicolumn{1}{|c@{\hspace{5.5mm}}|}{{\bf Example}} \\
\hline}
\tabletail{%
\hline
\multicolumn{5}{|r|}{\small\sl continued on next page}\\
\hline}
\tablelasttail{\hline}
\topcaption{New co-spherical elements while tessellating   1-irregular cubes}
\begin{supertabular}{| c@{\hspace{4mm}} | c@{\hspace{5.5mm}} | c@{\hspace{5mm}} | c@{\hspace{5mm}} | c@{\hspace{5.5mm}} |}
\shrinkheight{20.0mm}
\label{tab:interelem}
Generic \#11 & 7 & 11 & 6 & \raisebox{-\totalheight}{\includegraphics[scale=0.125]{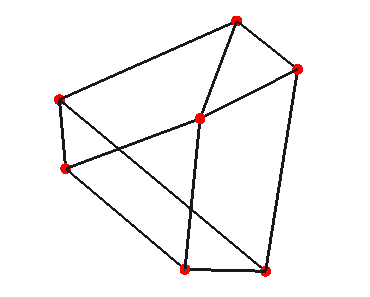}} \\ \hline
Generic \#12 & 7 & 11 & 6 & \raisebox{-\totalheight}{\includegraphics[scale=0.125]{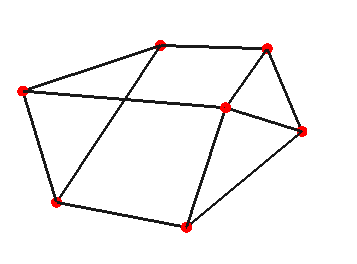}} \\ \hline
Generic \#13 & 7 & 12 & 7 & \raisebox{-\totalheight}{\includegraphics[scale=0.125]{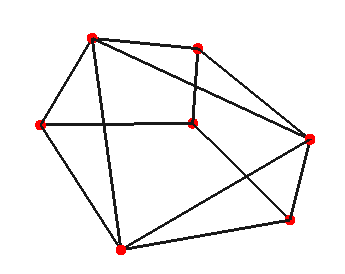}} \\ \hline
Generic \#14 &  7 & 13 & 8 & \raisebox{-\totalheight}{\includegraphics[scale=0.125]{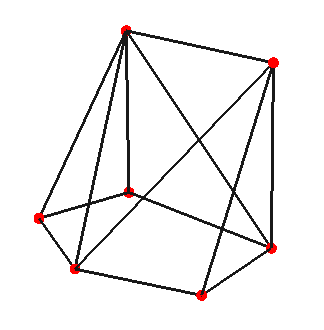}} \\ \hline
Generic \#15 & 7 & 14 & 9 & \raisebox{-\totalheight}{\includegraphics[scale=0.125]{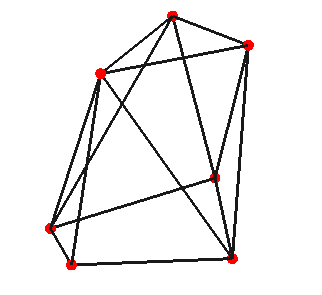}} \\ \hline
Generic \#16 & 8 & 12 & 6 & \raisebox{-\totalheight}{\includegraphics[scale=0.125]{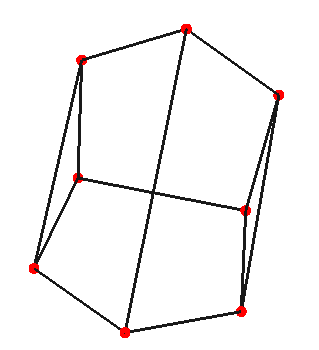}} \\ \hline
Generic \#17 & 8 & 12 & 6 & \raisebox{-\totalheight}{\includegraphics[scale=0.125]{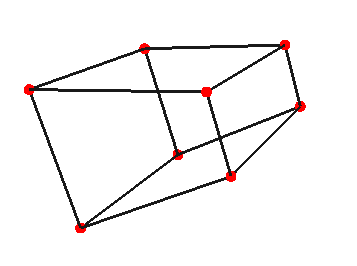}} \\ \hline
Generic \#18 & 8 & 13 & 7 & \raisebox{-\totalheight}{\includegraphics[scale=0.125]{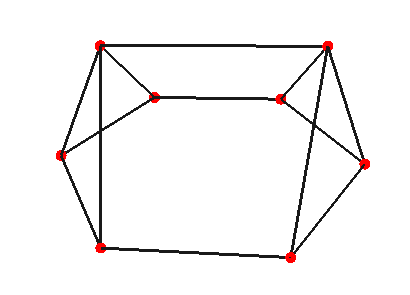}} \\ \hline
Generic \#19 & 8 & 14 & 8 & \raisebox{-\totalheight}{\includegraphics[scale=0.125]{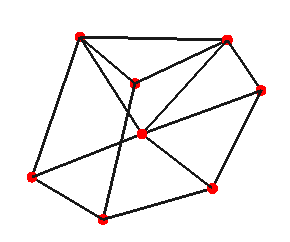}} \\ \hline
Generic \#20 & 8 & 15 & 9 & \raisebox{-\totalheight}{\includegraphics[scale=0.125]{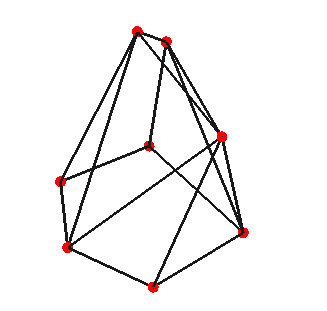}} \\ \hline
Generic \#21 & 9 & 15 & 8 & \raisebox{-\totalheight}{\includegraphics[scale=0.125]{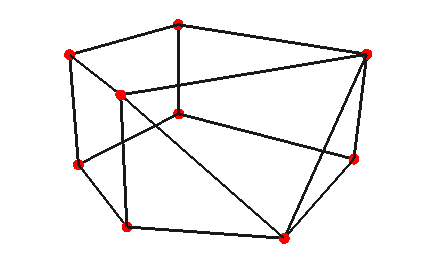}} \\ \hline
Generic \#22 & 9 & 15 & 8 & \raisebox{-\totalheight}{\includegraphics[scale=0.125]{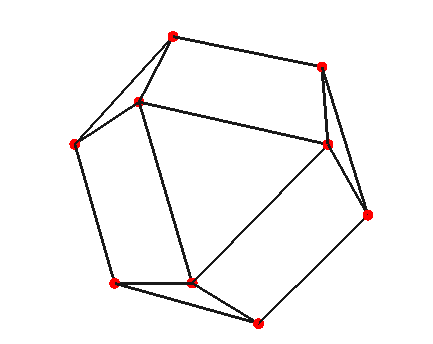}} \\ \hline
Generic \#23 & 9 & 16 & 9 & \raisebox{-\totalheight}{\includegraphics[scale=0.125]{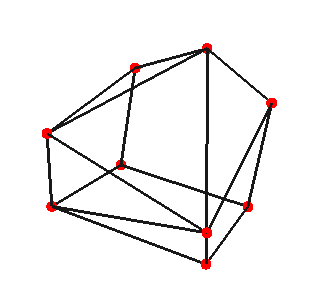}} \\ \hline
Generic \#24 & 9 & 16 & 9 & \raisebox{-\totalheight}{\includegraphics[scale=0.125]{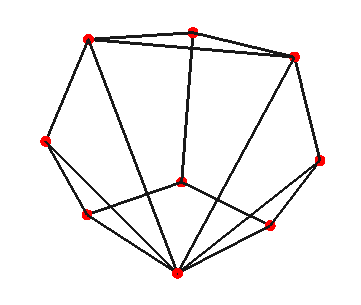}} \\ \hline
\end{supertabular}
\end{center}

\subsection{Evaluating the impact of each co-spherical element}

Table~\ref{tablaICO} shows a summary of the results obtained by generating the tessellations of all 1-irregular configurations 
of a cube:\\ \bigskip
\begin{center}
\tablefirsthead{%
\hline
\multicolumn{1}{|c@{\hspace{5mm}}}{{\bf Element}} &
\multicolumn{1}{|c@{\hspace{5.5mm}}||}{{\bf Freq.}} &
\multicolumn{1}{|c@{\hspace{5mm}}}{{\bf Element}} &
\multicolumn{1}{|c@{\hspace{5.5mm}}|}{{\bf Freq.}} \\
\hline}
\tablehead{%
\hline
\multicolumn{4}{|l|}{\small\sl continued from previous page}\\
\hline
\multicolumn{1}{|c@{\hspace{5mm}}}{{\bf Element}} &
\multicolumn{1}{|c@{\hspace{5.5mm}}||}{{\bf Freq.}} &
\multicolumn{1}{|c@{\hspace{5mm}}}{{\bf Element}} &
\multicolumn{1}{|c@{\hspace{5.5mm}}|}{{\bf Freq.}} \\
\hline}
\tabletail{%
\hline
\multicolumn{4}{|r|}{\small\sl continued on next page}\\
\hline}
\tablelasttail{}
\topcaption{Frequency of co-spherical elements while tessellating 1-irregular cubes}\label{tablaICO}
\begin{supertabular}{| l@{\hspace{5mm}} | r@{\hspace{5.5mm}} || l@{\hspace{5mm}} | r@{\hspace{5.5mm}} | }
     Cuboid &  531 &					Generic \#6 &  58 \\
     Tetrahedron &  39,590,100 &      		Generic \#7 &  1,881 \\ 
     Quadrilateral Pyramid &  5,200,926 &      	Generic \#8 &  2,340 \\ 
     Triangular Prism  &  184,374 &      		Generic \#9 &  6 \\ 
     Tetrahedron Comp. &  11,220 &     	 	Generic \#10 &  108 \\ 
     Def. Prism &  84,200 &      			Generic \#11 &  6,236 \\ 
     Def. Tetrahedron Comp. &  14,701 &      	Generic \#12 &  9,288 \\ 
     Pentagonal Pyramid  &  171,838 &      	Generic \#13  &  3,972 \\ 
     Hexagonal Pyramid  &  7,353 &      		Generic \#14 &  874 \\ 
     Triangular Bipyramid  &  625,447 &    	Generic \#15 &  4,966 \\ 
     Quadrilateral Bipyramid &  139,851 &      	Generic \#16 &  146 \\ 
     Pentagonal Bipyramid  &  25,686 &      	Generic \#17 &  204 \\ 
     Hexagonal Bipyramid  &  1,755 &      		Generic \#18 &  1,361 \\ 
     Biprism &  148 &      				Generic \#19 &  4,033 \\ 
     Generic \#1 &  61,044 &      			Generic \#20 &  197 \\ 
     Generic \#2 &  186,594 &      			Generic \#21 &  42 \\ 
     Generic \#3 &  94,020 &      			Generic \#22 &  214 \\ 
     Generic \#4 &  28,218 &      			Generic \#23 &  4 \\ 
     Generic \#5 &  28,028 &      			Generic \#24 &  6 \\ \hline
\multicolumn{1}{c}{} & \multicolumn{1}{c}{} & \multicolumn{1}{|l}{{\bf Total}} & \multicolumn{1}{|r|}{{\bf 46,491,970}}  \\
\cline{3-4}
\end{supertabular}
\end{center}
The trend observed in the bisection based approach is also observed here:
the most used elements are tetrahedra and quadrilateral pyramids ($\sim$85.15\% and $\sim$11.19\% of total elements respectively, corresponding to more than 96\%). The initial set of seven elements represents a $\sim$96.98\%, while the set of 24 elements found in configurations under the bisection based approach covers a $\sim$99.93\%. Finally, the elements that appear exclusively under the intersection based approach represent only a $\sim$0.07\% of the total.

\section{Conclusions}
\label{conclusiones}
We have identified 24 co-spherical elements  while
tessellating 1-irregular cubes generated by a bisection based approach
and 38 co-spherical elements while tessellating 1-irregular cubes generated by an intersection 
based approach.  We have experimentally noticed that in the tessellation of
1-irregular cubes (aspect ratio equal to 1) more co-spherical elements 
appear than in the tessellation of 1-irregular cuboids with larger aspect ratio. When we increase the 
cuboid aspect ratio a subset of these co-spherical
elements is required and no new co-spherical element appears. \\
\noindent We have studied the tessellations of 1-irregular cuboids generated
by a bisection based approach with three
different aspect ratios: 1, $\sqrt{2}$, and 4. The results can be summarized
as follows:\\
\begin{itemize}
\item All the tessellations for 1-irregular cubes and 1-irregular cuboids
with aspect ratio from 1  to $\sqrt{2}$ are adequate 
for the finite volume method. We would need to add 6 co-spherical elements to the
initial final element set if we want that our mixed element mesh generator can
tessellate the 1-irregular cuboids the first time the mesh is done 1-irregular.  
\item The number of different co-spherical elements while tessellating 
1-irregular cubes can be reduced from
24 to 16 by adding internal faces and to 13 by adding internal faces and
edges.   While tessellating 1-irregular cuboids with aspect ratio equal to
$\sqrt{2}$, the required elements are  reduced from 10 to 7 if we allow 
the insertion of internal faces. While tessellating 1-irregular cuboids with 
aspect ratio equal to 4  only 6 co-spherical elements are used.
\end{itemize}
 We have also study the tessellations of 1-irregular cubes generated
by an intersection based approach and 14 additional co-spherical elements 
appear. They represent less than 0.07\% of the total, then it  not useful to  include them in the set of final elements. They would increase this set  in $\sim$58\% (24 to 38). \\
\noindent It is worth to point out that the proposed algorithm was only applied for tessellating  1-irregular 
cuboids but it can also be used without any modification to tessellate
any 1-irregular convex configuration: 1-irregular prisms, pyramids or tetrahedra,
among others. Moreover, the algorithm can be used to generate Delaunay tessellations for any point set. It may be only required to recognize new co-spherical configurations. This means  we could apply this algorithm to the points of
a larger part of the 1-irregular mixed element mesh and not only to the 1-irregular basic elements. The circumcenter requirement is only
really necessary for 1-irregular elements that are located at the boundary or at a material interface.\\
\noindent We have made the study under the assumption that all the 1-irregular 
configurations
appear in the same rate, but this is for sure not true. While generating a mesh
based on modified octrees, there are some configurations that appear more
frequently than others. This fact could mean that some co-spherical elements
belonging to the tessellation of few 1-irregular cuboids, could have a 
greater impact than the one we have computed if these few configurations appear very frequently
while generating a mesh.  A complete study should consider also this case. \\ 
\noindent The study presented here is very useful for our mesh generator based
on modified octrees, but we think that it can also be useful for
other mesh generator based on octrees. 
\section*{Acknowledgments}

{\small
This work was supported by Fondecyt project 1120495. }

%
\bibliographystyle{ieeetr}
\bibliography{biblio}
%


\end{document}